\documentstyle[prd,aps,preprint,axodraw,12pt]{revtex}
\tightenlines
\input{epsf.sty}
\input{psfig.sty}

\newcommand{\be}{\begin{equation}}
\newcommand{\ee}{\end{equation}}
\newcommand{\bea}{\begin{eqnarray}}
\newcommand{\beas}{\begin{eqnarray*}}
\newcommand{\eea}{\end{eqnarray}}
\newcommand{\eeas}{\end{eqnarray*}} 
\newcommand{\ba}{\begin{array}}
\newcommand{\ea}{\end{array}}
\begin{document}

\draft


\title{ Lepton Masses from a TeV Scale in a 3-3-1 Model }

\author{J. C. Montero\footnote{e-mail:montero@ift.unesp.br},
C. A. de S. Pires\footnote{e-mail:cpires@ift.unesp.br}, 
V. Pleitez\footnote{e-mail:vicente@ift.unesp.br}}
\address{
  Instituto de F\'\i sica Te\'orica,
Universidade Estadual Paulista.\\
Rua Pamplona 145, 01405-900- S\~ao Paulo, SP
Brazil.}

\date{\today}

\maketitle
\begin{abstract}
In this work, using the fact that in 3-3-1 models the same leptonic bilinear
contributes to the masses of both charged leptons and neutrinos, we develop an
effective operator mechanism to generate mass for all leptons. 
The effective operators have dimension five
for the case of charged leptons and dimension seven for neutrinos.  
By adding extra scalar multiplets and imposing the discrete symmetry
$Z_9\otimes Z_2$ we are able to generate realistic textures for the leptonic 
mixing matrix. This mechanism requires new physics at the TeV scale. 

\end{abstract}


\section{Introduction}
\label{sec:intro}

The smallness of the neutrino masses and the pattern of their mixing, 
arising from atmospheric and solar neutrino data~\cite{atmos,soln,sno}, 
suggest the extension of the standard model. 
According to those experimental data, the mixing involved in
the atmospheric neutrino oscillation is maximal and the mixing 
involved in the solar neutrino oscillation is large~\cite{rev,sknew}. 
From the theoretical  point of view, we dispose 
already of well established ways where explanations to the smallness of those 
masses arise naturally. The most popular are the 
see-saw~\cite{seesaw} and the radiative generation~\cite{babu-zee} mechanisms.
Both of them require realistic extensions of the standard model and
extra global and/or discrete symmetries.
We also can consider effective operators which naturally lead to light
neutrinos. This approach has had success in accounting for the neutrino puzzle 
in the base of the standard model without resort to drastic fine tuning.

It was Weinberg~\cite{weinberg}, and independently Zee and 
Wilczek~\cite{wilczek}, who first pointed out in the context of the standard 
model that the dimension-five effective operator:
\begin{equation}
\frac{1}{\Lambda}\overline{L^c_{ia}}\,L_{jb}\varphi^{(m)}_k\varphi^{
(n)}_l
(f_{abmn} \epsilon_{ik}\epsilon_{jl}+
f^\prime_{abmn} \epsilon_{ij}\epsilon_{kl}),
\label{e1}
\end{equation}
with $L=(\nu_l,\, l)^T_L$, yields naturally light neutrino masses.
The success of such effective operator approach is justified by the expression of
the neutrino mass it generates:
\be
M^\nu_{ ab}= \frac{f_{ab} }{\Lambda}\,\langle \varphi \rangle^2,
\label{seesaw}
\ee 
which is a see-saw relation since $\langle\varphi\rangle\approx246$ GeV and 
$\Lambda$ is a large characteristic mass. Particularly, in this case the
realization of such operator turns to be important once it can guide us to all
the possible realizations of that mechanism using only the representation
content of the standard model with operators of dimension five~\cite{ema}.
Higher dimension operators have already been considered~\cite{babu}, and
extensions of the scalar sector of the standard model have also been
suggested~\cite{escext}. 

In this work we address the problem of generating neutrino and charged lepton
masses through effective operators in the context of 3-3-1 
models~\cite{331,scalsect}. In general, in these models, the minimal set of 
scalar multiplets required to generate the fermion masses consists in three 
scalar triplets: 
$\eta =( 
\eta^0 , \eta^-_1 ,  \eta^+_2)^T 
\sim({\bf 1},{\bf 3}, 0)$,   $\rho = (\rho^+, \rho^0 , \rho^{++})^T 
\sim({\bf 1},{\bf 3},1)$,  $\chi = (\chi^- , \chi^{--} , \chi^0)^T 
\sim({\bf 1},{\bf 3},-1)$ and  a symmetric sextet, $S \sim({\bf 1},{\bf 
6},0)$. 
An important feature of this model is that the same leptonic bilinear
$\overline{\Psi^c_L}\Psi_L$, where $\Psi_L=(\nu,\,l,\,l^c)^T$, can give a
contribution to the masses of both sectors: charged leptons and neutrinos. We
explore this feature in order to obtain realistic textures for the mixing 
matrix in the lepton sector. We work firstly in a scenario where there are only
triplets, like $\eta,\rho$ and $\chi$, and neutral scalar singlets. 
In this context, depending on the extra symmetry added to the model we
obtain realistic texture for the lepton masses. 
Next we show that a possible realization of this mechanism is
one in which heavy scalar sextets and singlet charged
leptons~\cite{dma,lepmass1} are added to the model.  
The scalar sextet which does not gain a vacuum expectation value induces a
dimension-five operator that give mass for the charged leptons; while the
dimension-seven operator, which give mass to the neutrinos, arises from the
interactions of neutrinos with the singlet charged leptons and a mixture
among the singly charged scalar bosons. 

The outline of this work is as follows. In Sec.~\ref{sec:II} we develop a 
formalism to address the issue of generating the charged lepton and neutrino
masses by effective operators in the context of a 3-3-1 model. Next, in
Sec.~\ref{sec:rs} we use the formalism previously developed to generate
realistic scenarios to accommodate lepton masses with appropriate mixing in the
lepton sector. In Sec.~\ref{sec:IV} we suggest what are the main ingredients a
more fundamental theory has to have in order to realize such effective
operators. We reserve the Sec.~\ref{sec:V} for our conclusions.  

\section{The mechanism}
\label{sec:II}

In this section we build effective operators of
dimension five for the generation of the charged lepton masses and dimension
seven for the case of neutrino masses. In particular we 
show that with a dimension-seven operator we attain the desired order of the 
neutrino masses required by the recent experiments, i.e., at the eV scale. 
It is interesting to note that the energy scale required to
obtain those neutrino masses is of the order of 5 TeV.

In the 3-3-1 models of Refs.~\cite{331,scalsect,dma} all leptons transform as
triplets under the electroweak gauge symmetry: 
\be
\Psi_{aL}=
\left (
\begin{array}{c}
\nu_a \\
l_a \\
l^c_a
\end{array}
\right )_L \,\, \sim (\mbox{{\bf 1}},\mbox{{\bf 3}},0);\quad a=e,\mu,\tau.
\label{leptrip} 
\ee

Below we generate the charged lepton masses by using the two scalar
triplets $\rho$ and $\chi$, mentioned in the previous section, and the
neutrino masses using the triplet $\eta$ and a neutral scalar singlet $\phi$. 

In order to generate realistic texture of the relevant matrices in the lepton
sector we will impose appropriate discrete or global symmetries (see the next
section). 
 
\subsection{Charged lepton masses}
\label{subsec:clmass}

In the version without the scalar sextet we still dispose of the 
scalar triplets $\rho$ and $\chi$ in order to generate the charged lepton
masses by dimension-five effective operators. 
According to the transformation properties under the symmetry $SU(3)_C
\otimes SU(3)_L \otimes U(1)_N $ of the last two scalar triplets we can form
with $\overline{\Psi^c_{aL}}\, \Psi_{bL}$ the following, 
effective dimension-five operator:   
\begin{eqnarray}
{\cal L} &=&\frac{f_{ab}}{\Lambda}\,\overline{\Psi^c_{aL}}\, \Psi_{bL} 
\chi^*\rho^* + H.c.\nonumber \\
&=&\frac{f_{ab}}{\Lambda}
\left\{ 
\overline{\nu^c_{aL}}
\left[ 
\nu_{bL} \chi^+ \rho^- + l_{bL} \chi^+ 
\rho^{0*} +  (l^c)_{bL} \chi^+ \rho^{--}
\right]
+\overline{(l_{aL})^c}
\left[ 
\nu_{bL} \chi^{++} \rho^- + 
 l_{bL} \chi^{++}\rho^{0*} \right.\right.\nonumber \\ &+&\left. \left. 
 l_{bL} \chi^{++} \rho^{--}
 \right]
+\overline{{l}_{aR}}
\left[ 
\nu_{bL} \chi^{0*} \rho^- +  l_{bL}
 \chi^{0*} \rho^{0*} + (l^c)_{bL} \chi^{0*} \rho^{--}
 \right]
 \right\} + H.c.
\label{5dimop}
\end{eqnarray}

After the neutral components $\chi_0$ and $\rho_0$ develop their respective 
VEVs, $\langle \chi \rangle $ and $\langle \rho \rangle$, the effective
operator above generates the following expression for the charged lepton mass 
matrix:
\be
M^l_{ab} = \frac{f_{ab}}{\Lambda} \langle \rho \rangle \langle \chi \rangle, 
\label{lepmassfor}
\ee
with $a,b=e,\mu,\tau$.
Let us discuss the values of the parameters present in the expression above. 
None of them have already been fixed by the model. We just expect that they 
can be found in some range of values. For example, $\langle \chi \rangle$ 
must be in the range: $ 300\, \mbox{GeV} < 
\langle \chi \rangle < 4\, \mbox{TeV} $\cite{ng,phf}.  The  constraint on
$\langle \rho \rangle$ comes from the mass of the gauge bosons  $W^{\pm}$ and
$Z^0$ i.e., $ \langle  \rho \rangle^2 +  \langle \eta 
\rangle^2 = (246)^2 $ GeV$^2$. Assuming, as an illustration, the
following  set of values:
\be
 \langle \eta \rangle \simeq 22\, {\rm GeV},\,\,\,\langle \rho
\rangle \simeq 245 \, \mbox{GeV},\,\,\,  \langle \chi \rangle \simeq 1
\,\mbox{TeV}, 
 \,\,\, \mbox{and}\,\,\,  
\Lambda \simeq 5\,\mbox{TeV},
\label{values}
\ee
the charged lepton mass matrix takes the form:
\be
M^l_{ab} \simeq 49 f_{ab}  \; \mbox{GeV}.  
\label{lepmassval}
\ee
If $f_{ab}$ is a diagonal matrix, for obtaining the correct charged lepton 
masses we need $f_{ee}\sim\times 10^{-5}$ for the electron mass, 
$f_{\mu\mu}\sim2\times10^{-3}$ for the muon mass, and $f_{\tau\tau}\sim3.6
\times10^{-2}$ for 
the tau mass. We recall that in the case of the standard model we have 
$10^{-6}$, $10^{-3}$ and $10^{-2}$, respectively.

\subsection{Neutrino mass}
\label{subsec:nusmass}

In order to generate neutrino masses we will consider only effective operators 
that conserve the total lepton number and also some other global or/and
discrete symmetries. The simplest way to obtain such operator is by adding 
an scalar singlet $\phi\sim({\bf1},{\bf1},0)$, coming from new physics at the
TeV scale, carrying the total lepton number, $L=L_e+L_\mu+L_\tau$, with the
following assignment $L(\phi)=-1$, and forming with $\eta$ and $\Psi_{aL}$ the 
dimension-seven effective operator:  
\bea
{\cal L} &=& \frac{ f^{\prime}_{ab} }{ \Lambda^3 }\overline{\Psi^c_{aL}}
\Psi_{bL} \eta^* \eta^* \phi \phi 
+H.c. \nonumber \\
&=&
\frac{ f^{\prime}_{ab} }{ \Lambda^3 }\left\{[\overline{\nu^c_{aL}}
\left[ \nu_{bL} \eta^{0*} \eta^{0*} +
l_{bL} \eta^{0*}\eta^+_1 + (l^c)_{bL} \eta^{0*} \eta^-_2 \right]
+ \overline{(l_{aL})^c}\left[ \nu_{bL} \eta^+_1 \eta^{0*} +
l_{bL} \eta^+_1 
\eta^+_1 +  (l^c)_{bL} \eta^+_1 \eta^-_2\right]\right. \nonumber \\
&+&\left. \overline{l_{aR}}\left[ \nu_{bL} \eta^{0*} \eta^-_2 + l_{bL} 
\eta^-_2 \eta^+_1 +  (l^c)_{bL} \eta^-_2 \eta^-_2 \right] \right\}\phi \phi
\nonumber \\ &+& H.c.
\label{7dimop}
\eea
Notice that this operator conserves the total lepton number, since
$L(\eta^+_2)=-2$ and $L(\eta^-_1)=L(\eta^0)=0$ (we recall that
$L(\chi^-)=L(\chi^{--})=2$ and hence the interactions in Eq.~(\ref{5dimop}) are
also $L$-conserving). The dimension-five operator
$\overline{\Psi^c_{aL}}\Psi_{bL} \eta^* \eta^*$ which violates the lepton number
explicitly can be forbidden by introducing discrete symmetries as we will show
below. For the moment we will avoid it. If we want this quantum number to be
broken spontaneously we should begin with $L$-conserving interactions and let a
non-zero VEV, in this case $\langle\phi\rangle$, to break this symmetry
spontaneously. A dangerous Majoron-like Goldstone can be avoided by breaking
softly or explicitly the total lepton number in the scalar potential (see
Eq.~(\ref{potential}) below) or by assuring that the Majoron is almost
singlet~\cite{bbd331}. Hence, after the scalars involved in Eq.~(\ref{7dimop})
develop their respective VEVs the neutrino masses are given by   
\be
M^\nu_{ab} = \frac{ f^{\prime}_{ab} }{ \Lambda^3 } \langle 
\eta \rangle^2 \langle \phi 
\rangle^2 .
\label{neutrexp}
\ee
Inserting  the values of the VEV given in Eq.~(\ref{values}) the expression
above reads 

\be
M^\nu_{ab} \simeq 3.9\, f^{\prime}_{ab}\left( \frac{\langle
\phi\rangle}{1\,{\rm GeV}}  \right)^2 \times 10^{-9}\,\mbox{GeV}. 
\label{neutrphi}
\ee

According to this the VEV involved above, $\langle \phi \rangle$, has to be 
around $5\times10^{-2}$ GeV if $f^\prime_{ab}\approx O(1)$, in order to 
generate the expected order of magnitude of the neutrino masses, that is, 
of the $10^{-2}$ eV order. 
That value of the VEV for the scalar singlet could imply a fine tuning since 
we expect the singlet scalar boson is very heavy, i.e., $m_\phi\approx\Lambda$.
However, in order to get such small VEV in a more natural way we can implement a
type II see-saw mechanism with the scalar field $\phi$~\cite{abdel}.

In fact, it is possible to implement this mechanism as we will
show in the following. Let us consider, for the sake of simplicity, 
the discrete symmetry $\eta \rightarrow-\eta$,  $\phi \rightarrow -\phi$ (other
fields are even under this symmetry). We also allow terms in the 
scalar potential that violate explicitly the total lepton number. In this case
the most complete scalar potential presenting these terms is:
\bea
V(\eta,\rho,\chi,\phi)&&=\mu^2_\eta\eta^{\dagger}\eta+\mu^2_\rho
\rho^{\dagger}\rho+
\mu^2_\chi\chi^{\dagger}\chi+ \mu_\phi \phi^{\dagger} \phi  
+\lambda_1(\eta^{\dagger}\eta)^2+\lambda_2(\rho^{\dagger}\rho)^2+
\lambda_3(\chi^{\dagger}\chi)^2 + \lambda_4 (\phi^{\dagger}\phi)^2 \nonumber \\
&&+(\eta^{\dagger}\eta)\left[ \lambda_5 (\rho^{\dagger}\rho) + 
\lambda_6(\chi^{\dagger}\chi)\right]+\lambda_7(\rho^{\dagger}\rho)(\chi^{
\dagger}\chi)+\lambda_8(\rho^{\dagger}\eta)(\eta^{\dagger}\rho)
+\lambda_9(\chi^{\dagger}\eta)(\eta^{\dagger}\chi)\nonumber \\ &&+
\lambda_{10}(\rho^{\dagger} 
\chi)(
\chi^{\dagger}\rho)
+(\phi^{\dagger} \phi) \left[ \lambda_{11} (\eta^{\dagger} \eta) 
+\lambda_{12} 
(\rho^{\dagger}\rho) + 
\lambda_{13}(\chi^{\dagger}\chi) \right]\nonumber \\
&&+[\lambda_{14}\epsilon \eta \rho \chi \phi + \lambda_{15}\chi^\dagger\eta
\rho^\dagger\eta+ H.c.],
\label{potential}
\eea
where the last two terms are those that violate explicitly the lepton number. 
From this scalar potential we find the following constraint equation
over $\langle \phi \rangle$ :
\be
\langle\phi\rangle[ \mu^2_\phi + \lambda_{11} \langle\eta\rangle^2 + 
\lambda_{12} \langle\rho\rangle^2 + 
\lambda_{13}\langle\chi\rangle^2 + \lambda_4 \langle\phi\rangle^2] +
 \lambda_{14}\langle\eta\rangle \langle \rho\rangle  \langle\chi\rangle = 
0.
\label{constr}
\ee
Supposing that $\mu^2_\phi<0$ is the dominant parameter in the term within
square brackets we have, 
\be
\langle \phi \rangle \simeq -\lambda_{14}\frac{\langle \eta \rangle \langle 
\rho \rangle  \langle \chi \rangle}{\mu^2_\phi}.
\label{seesawII}
\ee
Using the values in Eq.~(\ref{values}) in  Eq.~(\ref{seesawII}), and assuming
$\vert\mu_\phi\vert\approx \Lambda$ we obtain $\langle \phi \rangle 
\simeq 5\times10^{-2}$ GeV, if $\lambda_{14}=0.25$.
We recall that it was already shown in literature that it is possible 
to have a heavy scalar with a respective small VEV~\cite{ema,bbd331} as in the
present case.  

From Eq.~(\ref{neutrphi}) we find that the neutrino mass matrix
is given by the following expression:
\be
 M^\nu_{ab} \simeq 10^{-2}f^\prime_{ab}\; \mbox{eV}. 
\label{neutrinomass}
\ee

We see from the discussion above that both, charged leptons and neutrinos, gain
mass through effective operators, as can be seen from Eqs.(\ref{lepmassfor}) and
(\ref{neutrphi}) or (\ref{lepmassval}) and (\ref{neutrinomass}). However, the
scale of neutrino masses relatively to the charged lepton masses arise as a
consequence of the dimension of the effective operator and of the VEV of the
scalar involved.  

\section{Realistic scenarios}
\label{sec:rs}

If we want to obtain a definite texture on the mass matrices
in Eqs.~(\ref{lepmassfor}) and (\ref{neutrexp}) we have to enlarge the number 
of scalar multiplets, for instance one triplet of the type $\rho$ and $\chi$ 
and a singlet $\phi$ for each generation. We denote them as
$\rho_{1,2,3},\chi_{1,2,3},\phi_{1,2,3}$. In this case we have  
\begin{equation}
{\cal L}=\frac{f^{ij}_{ab}}{\Lambda}\,\overline{\Psi^c_{aL}}\, \Psi_{bL} 
\chi^*_i\rho^*_j+ \frac{ f^{\prime ij}_{ab} }{ \Lambda^3}
\overline{\Psi^c_{aL}} \Psi_{bL} \eta^* \eta^* \phi_i \phi_j 
+H.c., 
\label{new}
\end{equation}
and we will impose discrete symmetries in order to obtain appropriate mass
matrices. For instance, consider the symmetry $Z_9\otimes Z_2$ with fields
transforming under the $Z_9$ factor as
\begin{eqnarray}
\Psi_e\to \omega_1\Psi_e,\;\;\Psi_\mu\to\omega_3\Psi_\mu,\;\;
\Psi_\tau \to \omega_2\Psi_\tau,\;\;
\rho_1 \to \omega_3\rho_1,\;\;\chi_1 \to\omega^{-1}_1\rho_1,\nonumber \\
\rho_2\to\omega^{-1}_3\rho_2,\;\;\chi_2\to \omega_0\chi_2,\;\;
\rho_3\to\omega_4\rho_3,\;\;\chi_3\to\omega^{-1}_4\chi_3,\;\;
\eta\to \omega_0\eta,\nonumber \\ \phi_1\to \omega^{-1}_1\phi_1,\;\;
\phi_2\to \omega^{-1}_3\phi_2,\;\;\phi_3\to\omega^{-1}_2\phi_3,\;\;
\label{z6z2}
\end{eqnarray}
with $\omega_k=e^{2\pi ik/9}, \,k=0,\cdots,4$ and under the $Z_2$ factor the
fields $\Psi_\mu,\rho_1,\chi_1,\phi_{2,3}$ are odd,
while the other ones are even. (This implies appropriate transformation in the
quark sector if the scalar multiplets also couple to quarks.) Notice that the
$Z_9$ symmetry forbids the interactions $\epsilon\overline{\Psi^c_{aL}}\,
\Psi_{bL}\eta $ and $\overline{\Psi^c_{aL}}\, \Psi_{bL}\eta^*\eta^*$. 

From the interactions in Eq.~(\ref{new}) and the
discrete symmetries in Eq.~(\ref{z6z2}) we obtain mass matrices which 
mix the second and the third generation in the charged lepton sector, and
the first and the second generation in the neutrino sector, i.e.,  
they are diagonalized by following orthogonal transformations
\begin{equation}
U^l_L\stackrel{\sim}{=}\left(\begin{array}{ccc}
1 &0 &0\\
0 &c_0 & s_0\\
0 &-s_0 &c_o\end{array}\right),\quad
U^\nu_L\stackrel{\sim}{=}\left(\begin{array}{ccc}
c'_o &s'_0 &0\\
-s'_0 &c'_0 & 0\\
0 & 0& 1 \end{array}\right),
\label{ma1}
\end{equation}
for the charged lepton and neutrinos, respectively. The Maki-Nakagawa-Sakata
(MNS) mixing matrix is of the form~\cite{barr}: 
\begin{eqnarray}
U=U^{l\dagger}_LU^\nu_L=\left(\begin{array}{ccc}
c'_0 &s'_0 &0\\
-c_0s'_0 &c_0c'_0 & s_0\\
s_0s'_0 & -s_0c'_0& c_0 \end{array}\right),
\label{ma2}
\end{eqnarray}
and we have omitted Dirac or Majorana $CP$ violating phases. Notice that
$U_{e3}$ is zero at the tree level, and it must arise from radiative
corrections. After taking into account these corrections we expect that $U\to
V$: 
\begin{eqnarray}
V\approx\left(\begin{array}{ccc}
c_\odot &s_\odot & V_{e3}\\
-c_{\rm atm}s_\odot &c_{\rm atm}c_\odot & -s_{\rm atm}\\
-s_{\rm atm}s_\odot & s_{\rm atm}c_\odot& c_{\rm atm} \end{array}\right)
\simeq \left(\begin{array}{ccc}
c_\odot &s_\odot & V_{e3}\\
-\frac{1}{\sqrt2}s_\odot & \frac{1}{\sqrt2}c_\odot & -\frac{1}{\sqrt2}\\
-\frac{1}{\sqrt2}s_\odot & \frac{1}{\sqrt2}c_\odot& \frac{1}{\sqrt2}
\end{array}\right). 
\label{matrices2}
\end{eqnarray}

In this context, the small value of $\vert V_{e3}\vert(<0.16)$~\cite{chooz} is
natural in the present model because it arise from radiative corrections
involving a term in the scalar potential which breaks softly those discrete
symmetries. For instance, in a 
generalization of the expression in Eq.~(\ref{potential}) a term like
$\chi^\dagger_1\chi_2$ breaks softly the $Z_9$ symmetry. Hence, radiative
corrections should induce a small value for this entry on the MNS matrix and
(small) correction to the angles in Eqs.(\ref{ma1}) and (\ref{ma2}). In this
case $(c,s)_0\to (c,s)_{\rm 
atm},(c',s')_0\to (c,s,)_\odot$, $U_{e3}(=0)\to V_{e3}(\leq0.16)$. 
Since in general we expect that radiative correction does not amplify the 
mixing angles in Eqs.~(\ref{ma1}) and (\ref{ma2}) then 
$(c,s)_0\simeq (c,s)_{\rm atm}$ and $(c',s')_0\simeq (c,s)_\odot$. 
Notice that if the interactions that contribute to a non-zero value for 
$V_{e3}$ may induce a complex value for this entry inducing in this way a 
CP violating phase~\cite{miura}. 

Another possible scenario, where there is a bi-maximal mixing among the
neutrinos but not in the MNS matrix, appears as a result of a new symmetry
$L^{\prime}$ (in the context of the standard model $L^\prime$ can be identified
with $L_e -L_\mu -L_\tau$~\cite{escext,lprime,bimax} but here it must be an
independent global symmetry). In this case we need only two sort of $\rho$ and
$\chi$ scalar triplets and two singlets. We assign 
$L'(\Psi_e)=-L'(\Psi_\mu)=-L'(\Psi_\tau)=1$,
$L'(\rho_1)=L'(\chi_1)=-L'(\rho_2)=-L'(\chi_2)=1$ and $L'(\eta)=L'(\phi)=0$.  
With this symmetry we obtain from Eq.~(\ref{new}) a general
mass matrix in the charged lepton sector, and the following neutrino mass 
matrix at the tree level:  
\begin{eqnarray}
M^\nu=  \left(\begin{array}{ccc} 0 & f^{\prime 12}_{e\mu} & f^{\prime
12}_{e\tau}\\ f^{\prime 12}_{e\mu} & 0 & 0 \\
f^{\prime 12}_{e\tau} & 0 & 0\end{array}\right)
\frac{\langle
\eta\rangle^2\langle\phi_1\rangle\langle\phi_2\rangle}{\Lambda^3}.
\label{m1}
\end{eqnarray}
with $f^{\prime 12}_{e\mu}\sim f^{\prime 12}_{e\tau}$.
In this case the neutrino mass matrix has the inverse hierarchy: $(m,-m,0)$
where $m\propto [(f^{\prime 12}_{e\mu})^2 +(f^{\prime 12}_{e\tau} )^2]^{1/2}$. 
The mass splitting has to be generated by radiative corrections, as we will
show below, and we have
$\vert m_3\vert\ll \vert m_1\vert \simeq\vert m_2\vert$. 
Assuming that $ f^{\prime 12}_{e\mu}= f^{\prime 12}_{e\tau}$ there is a 
bi-maximal neutrino mixing pattern~\cite{lprime}:
\begin{eqnarray}
U^\nu= \left(\begin{array}{ccc} \frac{1}{\sqrt{2}} & -\frac{1}{\sqrt{2}}
& 0 \\ \frac{1}{2} & \frac{1}{2} & -\frac{1}{\sqrt{2}} \\ 
\frac{1}{2} & \frac{1}{2} & \frac{1}{\sqrt{2}} \end{array}\right). 
\end{eqnarray}

Since the bi-maximal mixing is not favored by the actual neutrino solar
data, i.e., $\tan~\theta_\odot~<1$~\cite{sknew}, we have to explain the
deviation from this bi-maximal scenario. 
This is possible to be done since from Eq.~(\ref{new}) and the $L'$ symmetry 
the mass matrix in the charged lepton sector, as we said before, is now a 
general one. Hence the respective left-handed mixing matrix $U^l$ can be 
written in terms of three angles $(c,s)_{12},(c,s)_{23},(c,s)_{13}$. 
In this case assuming that $s_{12}\gg s_{23} \gg s_{13}$ the mixing matrix 
$U^l$ is a hierarchical one similar to the mixing matrix in the quark sector 
and the charged lepton mass matrix is almost diagonal. Since the MNS matrix 
is defined as $V=U^{l\dagger}U^\nu$ the analysis of Ref.~\cite{giunti} 
follows.  

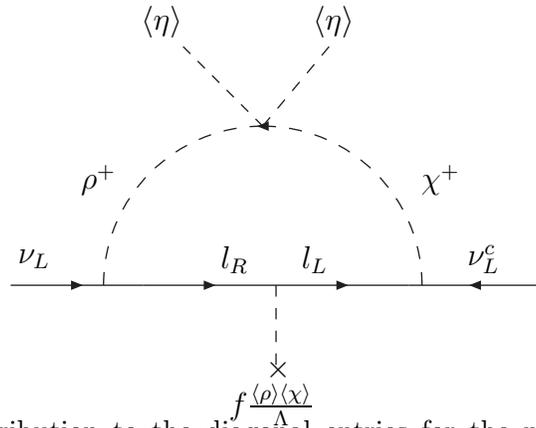
\begin{figure}
\vskip 0.6in  
\centerline{
\begin{picture}(300,300)(0,0)
\ArrowLine(50,50)(100,50)
\ArrowLine(100,50)(150,50)
\ArrowLine(150,50)(200,50)
\ArrowLine(250,50)(200,50)
\DashArrowArc(145,50)(60,0,180)5
\DashLine(150,50)(150,20){4}
\DashLine(145,110)(115,140){4}
\DashLine(145,110)(170,140){4}
\Text(90,90)[r]{$ \rho^+ $} 
\Text(220,90)[r]{$\chi^+ $} 
\Text(115,150)[r]{$ \langle \eta \rangle $} 
\Text(180,150)[r]{$\langle \eta \rangle $}
\Text(165,5)[r]{$f\frac{\langle \rho \rangle \langle \chi \rangle}{\Lambda} $}
\Text(65,60)[r]{$\nu_L$}  \Text(235,60)[r]{$\nu^c_L$}
\Text(140,60)[r]{$l_R$} \Text(170,60)[r]{$l_L$} 
\Text(156,18)[r]{$\times$}
\end{picture}
 } \caption{ One loop contribution to the diagonal entries for the neutrino 
mass matrix in the almost bi-maximal scenario.} 
\label{fig1}
\end{figure}

In this almost bi-maximal scenario we should address the mass splitting between
$\nu_1$ and $\nu_{2}$. For getting such a mass splitting we have to consider
terms in the scalar potential that break explicitly $L$, as is the case of the
last $\lambda_{15}$ term in the scalar potential in Eq.~(\ref{potential}). 
However, with the symmetry $Z_9$ introduced above this term involving
$\rho_i,\chi_i$ and $\eta$ is forbidden. Hence, it should be necessary that
$\eta$ transforms non-trivially under $Z_9$ (or a higher discrete symmetry) or,
only as an illustration, we can add a fourth pair of triplets $\rho_4$ and
$\chi_4$ which transform like $\eta$ (and for this reason they do not couple
directly with leptons) and with $L'(\chi_4)=L'(\rho_4)=0$. Thus we have the 
term $\lambda_{15}(\chi^\dagger_4\eta)(\eta^\dagger\rho_4)$. The scalar 
potential produces a general mixing among all the scalar of the same charge 
and we have interactions like $\chi^+\rho^-\langle \eta\rangle^2$, where 
$\chi^+,\rho^-$ denote symmetry eigenstates. That term, together with the 
interaction in Eq.~(\ref{new}), will generate corrections, through the one 
loop diagrams, for example to the diagonal entries in the mass matrix in 
Eq.~(\ref{m1}), providing the mass splitting between $\nu_1$ and $\nu_2$. 
The loop diagram which will generate, the diagonal entries in Eq.~(\ref{m1}) 
is depicted in Fig.~\ref{fig1}. It gives, up to logarithmic corrections, the 
following expression for such entries: 
\be
M^\nu_{aa}\simeq\frac{ \lambda_{15}f^3_{aa} \langle\eta\rangle^2
\langle\rho\rangle^2 
\langle\chi\rangle^2}{m^2_\chi \Lambda^3}.
\label{loopcorec}
\ee
All the parameters above, but $m_\chi$ and $\lambda_{15}$,  were
already previously fixed in this work. Assuming now that 
$m_\chi = \langle \chi \rangle$, and inserting Eq.~(\ref{values}) in
Eq.~(\ref{loopcorec}), we have: 
\be
M^\nu_{aa} \simeq 1.1\times 10^{-8}\lambda_{15} f^3_{aa}\, \mbox{GeV}.
\label{diagentry}
\ee

The large mixing angle MSW solution to the solar neutrino problem requires
a neutrino mass scale of the order of $2.8\times10^{-3}$ eV~\cite{gefan}. The
only free parameter in Eq.~(\ref{diagentry}) is $\lambda_{15}$, while the
diagonal $f$ parameters are already fixed by the charged lepton masses and 
their values are given in Sec.~\ref{subsec:clmass} (although in this case 
each entry of the matrix $f_{ab}$ has several contributions that we have not 
written explicitly).  We obtain this mass scale for the neutrinos by 
considering $\lambda_{15}\sim2.5\times10^{-4}$. Hence to generate the 
$\nu_1$-$\nu_2$ mass splitting we must to fine tuning $\lambda_{15}$. 

In the next section we analyze what main ingredients an
underlying theory should have to realize, in an economical way, the effective
operators and generate the two mixing scenarios considered above.

\section{A possible underlying theory}
\label{sec:IV}

The minimal scenario we can imagine is the one where the effective
dimension-five operator in Eq.~(\ref{5dimop}) is realized at the tree level,
while the effective dimension-seven operator in Eq.~(\ref{7dimop}) is
realized through the one loop level. 
For this we only need to add to the minimal 3-3-1 model (three quark generations
and three scalar triplets $\eta,\rho$ and $\chi$) 
at least two sextets $S^p,\,p=1,2$ which do not necessarily gain a non-zero VEV, 
two heavy lepton singlets, $E_{1,2(L,R)} \sim({\bf 1},{\bf 1},
-1)$~\cite{dma,lepmass1} and three scalar singlets $\phi^r,\,r=1,2,3$ like the
$\phi$ introduced in Sec.~\ref{subsec:nusmass}. 
We will assume that these singlet leptons have the following assignments of 
the total lepton number: $L(E_1)=1$ and $L(E_2)=0$. 
We recall that it was shown in Ref.~\cite{effop} that a tree level realization
of a symmetric bilinear $\overline{\Psi^c_L}\Psi_L$ is implemented by
introducing a scalar sextet. 

\begin{figure}
\vskip 0.6in  
\centerline{
\begin{picture}(300,300)(0,0)
\ArrowLine(50,200)(155,200)
\ArrowLine(148,200)(260,200)
\DashLine(155,200)(155,260){4}
\DashLine(155,260)(100,300){4}
\DashLine(155,260)(200,300){4}
\Text(175,230)[r]{$S$} 
\Text(105,315)[r]{$\langle \rho \rangle$} 
\Text(230,315)[r]{$\langle \chi \rangle$}  
\Text(80,185)[r]{$l_L$}  \Text(240,185)[r]{$l_R$}
\Text(40,200)[r]{a)}
\ArrowLine(50,50)(250,50) 
\DashArrowArc(145,50)(60,0,180)5
\DashLine(120,50)(120,20){4}
\DashLine(180,50)(180,20){4}
\DashLine(145,110)(115,140){4}
\DashLine(145,110)(170,140){4}
\Text(90,90)[r]{$ \rho^+ $} 
\Text(220,90)[r]{$\chi^+ $} 
\Text(115,150)[r]{$ \langle \eta \rangle $} 
\Text(180,150)[r]{$\langle \eta \rangle $}
\Text(130,10)[r]{$ \langle \phi \rangle $} 
\Text(190,10)[r]{$\langle \phi \rangle $}
\Text(65,60)[r]{$\nu_L$}  \Text(235,60)[r]{$\nu^c_L$}
\Text(110,60)[r]{$E_{1R}$} \Text(145,60)[r]{$E_{2L}$} \Text(155,40)[r]{$M_2$}
\Text(176,60)[r]{$E_{2R}$} \Text(200,60)[r]{$E_{1L}$}
\Text(155,50)[r]{\large $\times$}  
\Text(40,50)[r]{b)}
\end{picture}
 } \caption{ Tree level and 1-loop diagrams contributing to the effective
 operators defined in Eqs.~(\ref{5dimop}) and (\ref{7dimop}), respectively.}   
 \label{fig2}
\end{figure}
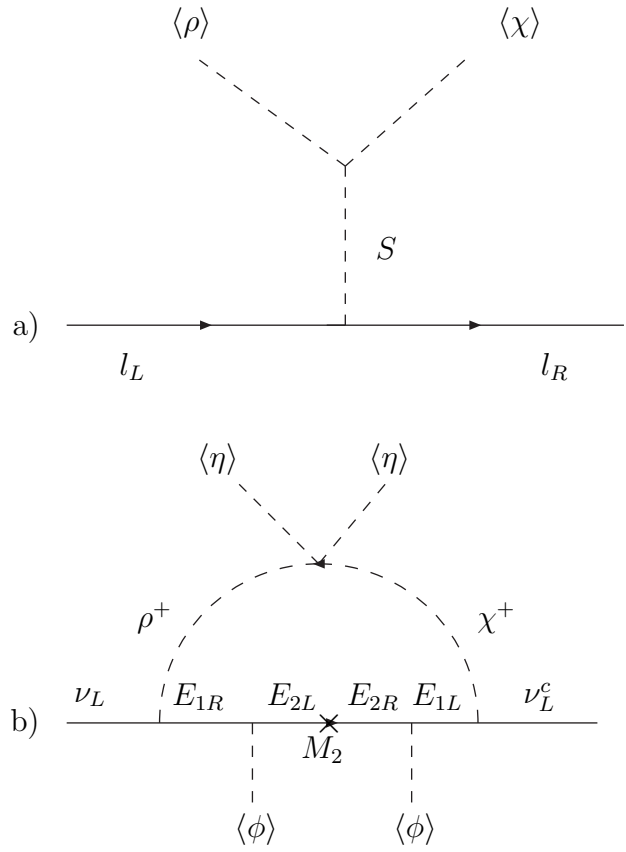

With the representation content discussed above we have the leptonic
interactions:     
\begin{eqnarray}
{\cal L}&=& G^p_{ab}\,\overline{(\Psi^c)_{aiL}}\Psi_{bjL}S^p_{ij}
+ G_{1a}\,\overline{\Psi_{aL}} E_{1R} \rho + 
G_{2a}\chi^{T}\,  \overline{
N_{1L}}\Psi^c_{aL}+g^\prime_r\,\overline{E_{1L}}E_{2R}\phi^{r}+
g_r\,\overline{E_{2L}}E_{1R}\phi^{r*} \nonumber \\ &+& 
M_1\overline{E_{1L}}E_{1R} 
+M_2\overline{E_{2L}} E_{2R}+ H.c., 
\label{undthe}
\end{eqnarray}
with $M_1 , M_2 \simeq \Lambda$ and we have omitted $SU(3)$ indices and
summation over the repeated indices. By imposing an appropriate discrete
symmetry we can get that one of the scalar sextets couples only to the first
leptonic generation and the other one to the second and third generations. 
Hence, with the interactions above we can realize the effective dimension-five
operator in Eq.~(\ref{new}) as is shown in the Fig.~\ref{fig2}a, where the
trilinear vertex arises from a term like $g\chi^T S^\dagger\rho$ where $g$ is a
constant with dimension of mass. So, we obtain a mass matrix for the charged
lepton which is diagonalized by an orthogonal matrix, like $U^l_L$ in
Eq.~(\ref{ma1}) and from Eq.~(\ref{5dimop}) we see that
$1/\Lambda=g/M^2_S$. For the realization of the effective dimension-seven
operator the last term of the scalar potential in Eq.~(\ref{potential}) is also
important. Such realization is depicted in Fig.~\ref{fig2}b.

\section{conclusions }
\label{sec:V}

In this work we developed a simple mechanism based on effective operators in 
the context of a 3-3-1 model  which generates masses for the neutrinos and 
for the charged leptons as well.  
By using the same bilinear the charged lepton masses are generated in this
mechanism by an effective dimension-five operator, while the neutrino masses
require an effective dimension-seven operator in conjunction with a type II 
see-saw mechanism applied on a scalar singlet.  
Then, we use the effective operator mechanism to generate 
several mixing matrices in the lepton sector which are consistent with the 
solar, atmospheric and reactor neutrino data. 
For this we have considered the discrete symmetry $Z_9\otimes Z_2$ 
or a global one, $L^{\prime}$, in the effective operators given in
Eq.~(\ref{new}). In this case, without resort to large fine tuning we obtain
neutrino masses compatible with some solutions to the solar and atmospheric
neutrino anomalies. We would like to stress that in this 3-3-1 model the
bilinear $\overline{\Psi^c}\Psi$ gives mass to both lepton sectors. 
This issue is a characteristic of the 3-3-1 models: in the standard model 
the bilinear $\overline{L^c}L$ can only contribute to the neutrino masses.   

Although we have not considered the quark masses we would like to call the 
attention to an interesting mechanism, in the context of a 3-3-1 model, for
generating the top and bottom masses at the tree level, while the masses of the
other quarks and charged leptons arise at the 1-loop level, which was proposed
in Ref.~\cite{mdt}. 

\acknowledgments 
This work was supported by Funda\c{c}\~ao de Amparo \`a Pesquisa
do Estado de S\~ao Paulo (FAPESP), Conselho Nacional de 
Ci\^encia e Tecnologia (CNPq) and by Programa de Apoio a
N\'ucleos de Excel\^encia (PRONEX).


%

\end{document}